\documentclass[RNAAS]{aastex62}

\begin{document}

\title{Proving Heliocentrism and Measuring the Astronomical Unit in a Laboratory Astronomy Class via the Aberration of Starlight}

\correspondingauthor{Jason T.\ Wright}
\email{astrowright@gmail.com}
\author[0000-0001-6160-5888]{Jason T.\ Wright}
\affil{Department of Astronomy \& Astrophysics \\ and \\ Center for Exoplanets and Habitable Worlds \\ 525 Davey Laboratory \\
The Pennsylvania State University \\
University Park, PA, 16802, USA}


\section{} 

The objective reality of the Earth's motion about the Sun was finally proven observationally by \citet{Bradley27} when he correctly explained the $\sim 20\arcsec$ annual, elliptical motions of stars as being due to aberration of starlight caused by the motion of the Earth.


The effect is so large that at first blush it seems that it should be quite easily measured with modern equipment of the sort used in astronomy laboratory courses, but since the effect is an {\it absolute} displacement of {\it all} stars, it is not detectable with the usual methods of {\it differential} astrometry. One must instead measure stellar positions with respect to some absolute reference direction. Happily, there is an absolute reference readily available to many telescopes: the true Celestial Poles, which are, by definition, directions parallel to the Earth's rotation axis.

To use this reference, one could point an imaging telescope at the Celestial Pole {\it and turn off tracking}, taking exposures throughout the night. Then, a stack of all of these images should reveal star trails centered on the true Celestial Pole, allowing one to measure its position with respect to the {\it apparent} positions of the stars. Since the stars (unlike the poles) are subject to aberration, they will appear to move throughout the year with respect to the Celestial Pole by $\sim 20\arcsec$.

Repeating this measurement throughout the year would show the  Celestial Pole to move with respect to the apparent positions of the stars in an ellipse with a phase and orientation consistent with heliocentric models of the Sun's apparent motion on the Celestial Sphere, thus proving that this motion is actually caused by the the Earth's orbit around the Sun.  The angular size of this ellipse measures the velocity of the Earth's motion in units of the speed of light. It is a common physics laboratory exercise to measure the speed of light, and the two experiments together constitute a measurement of the astronomical unit.

There are some complications to consider when making such a measurement.

Many telescopes (or the buildings they sit upon) may be subject to significant swaying from wind. Many polar-mounted telescopes have Cassegrain-mounted cameras have northern declination limits. The site, aperture, and camera will need to permit the detection of trails of stars fainter than Gaia $G \sim 15$  within a few arcminutes of the true Celestial Pole.

The image and data analysis will also be complex. 

The aberration must be separated from the motion of the pole due to axial precession and nutation (a $\sim 20 \arcsec$ yr$^{-1}$ effect), and measuring all three effects is a multi-year project. Further, determining the centers of the star trails is not a standard image analysis problem, and differential refraction may mean that the star trails do not trace perfect circles (alternating exposures between two filters may allow one to measure and back out this effect). 

Fortunately, Polaris sits $\sim 45\arcmin$ from the North Celestial Pole, and so should not interfere with the measurement for Northern observers. The rotational and eccentric orbital motion of the Earth is smaller than its circular orbital motion by two orders of magnitude, so can both be ignored, as can polar motion, which is a $< 1\arcsec$ effect. The Celestial Poles are at $|b| = 27\arcdeg$, and so finding nearby stars should not pose too large a problem; there are multiple stars with $g<17$ within $2\arcmin$ of the North Celestial Pole for the next several years at least.

\begin{figure}
\hspace{-2.5cm}
\includegraphics[width=9in]{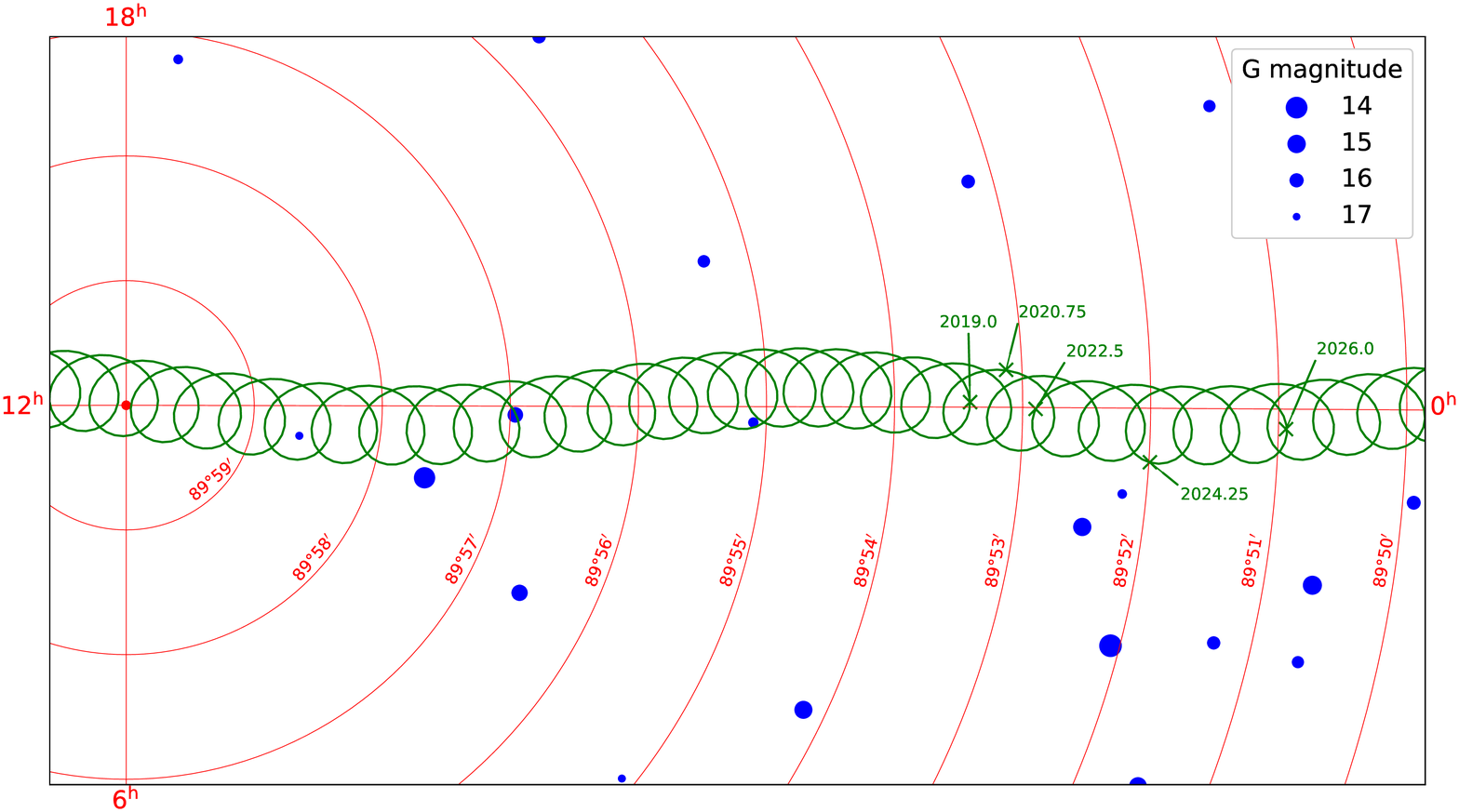}
\caption{Path, in green, of the apparent position of the North Celestial Pole (i.e.\ the center of star trails) as a function of time in ICRS coordinates. For reference, a few dates are identified in Julian years.  Most of the secular motion along RA=0$^{\rm h}$ is due to the precession of the Earth's axis; the rest, as well as the slow 18.6 year east-west variation, is due to nutation. The annual ``loops'' are caused by the aberration of the positions of the {\it stars} (shown here fixed at their ICRS positions from Gaia DR2). Detection of the loops constitutes observational proof of Earth's orbital motion, and measurement of their amplitude constitutes a measurement of the astronomical unit.}
\end{figure}

Despite my aspirations, I have not found the time to even begin the project. I nonetheless feel that it is a rich project in observational astronomy and fundamental physics, and a pleasingly didactic exercise appropriate for upper-division undergraduate and graduate astronomy and physics laboratory classes. True, it cannot be completed in a single semester, but its complexity and length may be justified by the novelty of proving heliocentrism with nothing but a small telescope and camera.

I therefore publish this Research Note to disseminate the idea in hopes that I might receive word in a few years of a triumphant astronomy class's proof of heliocentrism, along with their best estimates of the astronomical unit and precession.

\acknowledgments 
This Research Note has made use of NASA's Astrophysics Data System; Astropy, a community-developed core Python package for astronomy \citep{astropy}; and Gaia \citep{GaiaMission} DR2 \citep{GaiaDR2}.

\bibliography{references}

\end{document}